\documentclass[twocolumn,showkeys,preprintnumbers,amsmath,amssymb]{revtex4}

\usepackage{graphicx}
\usepackage{dcolumn}
\usepackage{bm}
\usepackage{epsf}

\begin{document}

\preprint{To appear in CHAOS, Volume 17, Issue 2, June 2007} 

\title{On the existence of chaotic circumferential waves in spinning disks}

\author{Arzhang Angoshtari}
 \email{arzhang_a@mehr.sharif.edu}
\author{Mir Abbas Jalali}
 \email{mjalali@sharif.edu}
 \homepage{http://sharif.edu/~mjalali}
\affiliation{
Center of Excellence in Design, Robotics and Automation \\
Department of Mechanical Engineering, Sharif University of
Technology \\ P.O.Box: 11365-9567, Azadi Avenue, Tehran, Iran }

\date{\today}

\begin{abstract}
We use a third-order perturbation theory and Melnikov's method
to prove the existence of chaos in spinning circular disks subject
to a lateral point load. We show that the emergence of transverse
homoclinic and heteroclinic points respectively lead to a random
reversal in the traveling direction of circumferential waves and
a random phase shift of magnitude $\pi$ for both forward and
backward wave components. These long-term phenomena occur in
imperfect low-speed disks sufficiently far from fundamental
resonances.
\end{abstract}


\maketitle

\label{sec:level1} {\bf Transversal vibration modes of hard disk
drives (HDDs) are excited by the lateral aerodynamic force of the
magnetic head. Previous works \cite{RM01,JA1} revealed that chaotic
orbits are inevitable ingredients of phase space flows when the
lateral force is large, or the disk is rotating near the critical
resonant speed. For low-speed disks, however, an adiabatic invariant
(a first integral) was found \cite{JA1} using a first-order
averaging based on canonical Lie transforms. According to the
first-order theory, regular vibrating modes of imperfect, low-speed
disks are independent of the angular velocity of the disk,
$\Omega_0$. In such a circumstance, the speed of circumferential
waves is the natural frequency of the lateral mode, $\omega$,
derived from linear vibration analysis. HDDs are usually operated
with angular velocities smaller than $\omega$ (safely below
resonance). Moreover, the magnitude of the lateral force $F$ is very
small. Given these conditions, we show that it is impossible to
continue the Lie perturbation scheme \cite{JA1} up to terms of
arbitrary order and remove the time variable $t$ from the
Hamiltonian. In fact, due to the special forms of nonlinearities in
the dynamical equations of spinning disks, one can not remove $t$
from third-order terms. Subsequent application of a second-order 
Melnikov theory reveals that transverse homoclinic and heteroclinic 
points do exist for all $F,\Omega_0\not =0$. This implies chaos, or
equivalently, non-integrability of governing equations}.

\section{Introduction}
\label{sec:intro} Dynamics of continuum media, like fluids, rods,
plates and shells is usually formulated as a system of partial
differential equations (PDEs) for physical quantities in terms of
the spatial coordinates $\textbf{x}$ and the time $t$ as
\begin{equation}
{\cal L}(\textbf{u})=0. \label{eq:field-equation}
\end{equation}
Here ${\cal L}$ is a nonlinear operator and
$\textbf{u}(\textbf{x},t)$ is the vector of dependent variables.
When the boundary conditions are somehow simple, approximate
variational methods based on modal decomposition and Galerkin's
projection \cite{Reddy86} can be used to reduce the order of
governing equations. These methods begin with solving an auxiliary
eigenvalue problem, which is usually the variational field equation
$\delta {\cal L}(\textbf{u},\Lambda)=0$, and build some complete
basis set $\textbf{U}_{\textbf{k}}(\textbf{x},\Lambda_{\textbf{k}})$
for expanding $\textbf{u}(\textbf{x},t)$ in the spatial domain. Here
$\Lambda_{\textbf{k}}$ is an {\it eigenvalue}, which is
characterized by the vectorial index $\textbf{k}$. Each
$\textbf{U}_{\textbf{k}}$ is called an eigenmode or a shape
function. Such a basis set should preferably satisfy boundary
conditions and be orthogonal.

Once a complete basis set is constructed, one may suppose a solution
of the form
\begin{equation}
\textbf{u}(\textbf{x},t)=\sum_{\textbf{k}}
\textbf{V}_{\textbf{k}}(t) \cdot
\textbf{U}_{\textbf{k}}(\textbf{x},\Lambda_{\textbf{k}}).
\label{eq:modal-expansion}
\end{equation}
Substituting from (\ref{eq:modal-expansion}) into
(\ref{eq:field-equation}) and taking the inner product
\begin{equation}
\int {\cal L}(\textbf{u})\cdot
\textbf{U}_{\textbf{k}'}(\textbf{x}){\rm d}\textbf{x}=0,
\end{equation}
leaves us with a system of nonlinear ordinary differential equations
(ODEs) for the amplitude functions $\textbf{V}_{\textbf{k}}(t)$. The
evolution of the reduced ODEs shows the interaction of different modes
and their influence on the development of spatiotemporal patterns.

In a series of papers, Raman and Mote \cite{RM01,RM99} used the
modal decomposition method to investigate transversal oscillations
of spinning disks whose deformation field is described in terms
of the displacement vector $(u,v,w)$ with $u$ and $v$ being the
in-plane components. The most important application of the spinning
disk problem is in design, fabrication and control of HDDs.
The governing PDEs for the
evolution of displacement components were first derived by Nowinski
\cite{N64} and reformulated more recently by Baddour and Zu
\cite{BZ01}. Let us define $(R,\phi)$ as the usual polar coordinates.
For a rotating disk with the angular velocity
$\Omega_0$, Nowinski's theory assumes that the in-plane inertias
$(u,v)\Omega_0^2$, $2\Omega_0(u_{,t},v_{,t})$ and
$(u_{,tt},v_{,tt})$ are ignorable against $R\Omega_0^2$. This is a
rough approximation for high-speed disks and one needs to use the
complete set of equations as Baddour and Zu \cite{BZ01} suggest.
Nowinski's theory, however, has its own advantages (like the
existence of a stress function) that facilitate the study of the
most important transversal modes. In low-speed disks, or disks with
high flexural rigidity, one has $\Omega_0\ll \omega$. Hence, it is
legitimate to apply Nowinski's governing equations in such systems.

In this paper we {\it analytically prove} the existence of chaos,
and therefore, non-integrability of the reduced ODEs that govern the
double-mode oscillations of imperfect spinning disks. We investigate
low-speed disks subject to a lateral point force exerted by the
magnetic head. The lateral force in HDDs is very small and its
origin is the aerodynamic force due to air flow in the gap between
the disk and the head. We show that chaotic circumferential waves
dominate some zones of the phase space over the time scale $t\sim
{\cal O}(\epsilon ^{-3})$ with $\epsilon$ being a small perturbation
parameter. This indicates very slow evolution of random patterns,
and the practical difficulties of their identification.

The paper is organized as follows. In \S\ref{sec:formulation}, we
present the Hamiltonian function in terms of Deprit's \cite{D91}
Lissajous variables. In \S\ref{sec:normalization}, we use a
canonical perturbation theory \cite{D69,DE91} to eliminate the fast
anomaly $l$ from the Hamiltonian. The action associated with $l$
then becomes an adiabatic invariant. Transversal intersections of
destroyed invariant manifolds, and therefore, non-integrability of
the normalized equations, is proved by a second-order Melnikov method 
in \S\ref{sec:melnikov}. We present a complete classification of
circumferential waves in \S\ref{sec:wave-classification} and
end up the paper with concluding remarks in \S\ref{sec:conclusions}.

\section{Problem formulation}
\label{sec:formulation} Let us assume $U_m(R)$ as an orthogonal
basis set that represents the disk deformation in the radial
direction. The index $m$ stands for the number of radial nodes that
$U_m(R)$ has. According to Raman and Mote's \cite{RM01} treatment of
imperfect disks, the following choice of the transversal displacement
field
\begin{equation}
w(R,\phi,t)=U_m(R)[x(t)\cos n\phi+y(t)\sin n\phi],
\label{eq:w-R-phi-t}
\end{equation}
reduces Nowinski's governing equations to a system of ODEs
for the amplitude functions $x(t)$ and $y(t)$
as
\begin{subequations}
\label{eq:reduced-ODE}
\begin{eqnarray}
\ddot x + \lambda^2 x +\epsilon \gamma \left ( x^2 +
    y^2 \right ) x \! &=& \! \epsilon F\cos
(n\Omega_0 t), \label{eq:x-equation} \\
\ddot y + \omega^2 y +\epsilon \gamma \left ( x^2 + y^2 \right ) y
\! &=& \! \epsilon F\sin (n\Omega_0 t), \label{eq:y-equation}
\end{eqnarray}
\end{subequations}
where $\lambda$, $\omega$ and $\gamma$ are constant parameters that
depend on the geometry and material of the disk. $\epsilon$ is
a small perturbation parameter, $F$ is the weighted integral of
the lateral point force, and $\Omega_0$ is the angular velocity of
the disk.

We suppose small deviations from perfect disks and write the
constant parameter of (\ref{eq:reduced-ODE}) as
$\lambda^2/\omega^2=1+\epsilon\eta$. We also define
$n\Omega_0=\epsilon \Omega$ with ${\cal O}(\Omega)\sim {\cal
O}(\omega)$. Denoting $(p_x,p_y)$ as the momenta conjugate to
$(x,y)$, it can be verified that equations (\ref{eq:reduced-ODE})
are derivable from the Hamiltonian function
\begin{eqnarray}
H &=& \frac 12 \left ( p_x^2 + p_y^2 \right )+
      \frac 12 \omega^2 \left ( x^2 + y^2 \right )
      + \epsilon \Big [ \frac {\eta}2 \omega ^2 x^2 \nonumber \\
  &+& \Omega P \!+\! \frac {\gamma}4 \left ( x^2 + y^2 \right )^2
      - F \left ( x \cos p + y \sin p \right ) \Big ] .
\label{eq:Hamiltonian-xy}
\end{eqnarray}
We have introduced the action $P$ and its conjugate angle
$p=\epsilon \Omega t$ to make our equations autonomous, which is a
preferred form for the application of canonical perturbation
theories. The {\it extended phase space} has now dimension six.
Dynamics generated by (\ref{eq:Hamiltonian-xy}) is better
understood after carrying out a canonical transformation
$(x,y,p_x,p_y)\rightarrow (l,g,L,G)$ to the space of Lissajous
variables \cite{D91} so that
\begin{subequations}
\label{eq:lissajous}
\begin{eqnarray}
x &=& s\cos(g+l)-d\cos(g-l), \\
y &=& s \sin (g+l)-d \sin (g-l), \\
p_x &=& -\omega \left [ s \sin (g+l)+ d \sin (g-l) \right], \\
p_y &=& \omega \left [ s \cos (g+l)+ d \cos (g-l) \right], \\
s &=& \sqrt{{L+G\over 2\omega}},~~d=\sqrt{{L-G\over 2\omega}},~~
L\ge 0,~~\vert G \vert \le L. \nonumber
\end{eqnarray}
\end{subequations}
In the space of Lissajous variables, the Hamiltonian
defined in (\ref{eq:Hamiltonian-xy}) becomes
\begin{eqnarray}
H &=& H_{0}(L) + \epsilon H_{1}(l,g,p,L,G,P),
\label{eq::Hamiltonian-Lissajous} \\
H_{0} &=& \omega L, \nonumber \\
H_{1} &=& \Omega P-F \left [ s \cos (g+l-p)-
                             d \cos (g-l-p)
                     \right ] \nonumber \\
&+&   {\gamma \over 4} \left [ (s^2+d^2)-2sd\cos(2l) \right ]^2
              \nonumber \\
&+&   {\eta \omega^2 \over 4} \Big [
(s^2+d^2)+s^2\cos(2g+2l)-2sd\cos (2g)
              \nonumber \\
&+&   d^2\cos(2g-2l)-2sd\cos (2l) \Big ]. \nonumber
\end{eqnarray}
From (\ref{eq::Hamiltonian-Lissajous}) we conclude that $l$ is the
fast angle, and $g$ and $p$ are the slow ones. Therefore the
long-term behavior of the flows generated by
(\ref{eq::Hamiltonian-Lissajous}) can be analyzed by averaging $H$
over $l$. After removing $l$, its corresponding action $L$ will be a
constant of motion for the flows generated by the averaged
Hamiltonian $\langle H \rangle _l$, and the phase space dimension
reduces from 6 to 4.

\section{Canonical third-order averaging}
\label{sec:normalization}
In order to average $H$ over $l$, we use the normalization procedure
of Deprit and Elipe \cite{DE91}. Denoting $X\equiv(l,g,p)$ and
$Y\equiv(L,G,P)$, we define a Lie transformation
$(l,g,p,L,G,P)\rightarrow(\bar l,\bar g,\bar p,\bar L,\bar G,\bar
P)$ as
\begin{equation}
X = E_{W}(\bar X),~~Y=E_{W}(\bar Y),\label{eq:Pri-unPri}
\end{equation}
so that the Hamiltonian function in terms of the new variables,
$K\equiv \langle H\rangle _l$, does not depend on $\bar l$. $E_{W}$
is the Lie transform generated by the function $W$ and it is defined
as
\begin{eqnarray}
E_{W}(\bar Z) &=& \bar Z+(\bar Z;W)+\frac{1}{2!}((\bar Z;W);W) \nonumber \\
&+& \frac{1}{3!}(((\bar Z;W);W);W)+\cdots . \label{eq::DifE_W}
\end{eqnarray}
In this equation, $(f_1;f_2)$ denotes the Poisson bracket of $f_{1}$
and $f_{2}$ over the $(\bar l,\bar g,\bar p,\bar L,\bar G,\bar
P)$-space. We expand the generating function $W$ as
\begin{equation}
W = \epsilon W_{1}+\frac{\epsilon^2}{2!}W_{2} + \cdots,
\label{eq::Gener_fun}
\end{equation}
and specify the averaged, {\it target Hamiltonian} $K=K(\bar g,\bar
p,\bar L,\bar G,\bar P)$ as the series \cite{DE91}
\begin{equation}
K = K_{0} + \epsilon K_{1} +
\frac{\epsilon^2}{2!}K_{2} + \frac{\epsilon^3}{3!}K_{3} + \cdots,
\label{eq::Kamiltonian}
\end{equation}
with
\begin{subequations}
\label{eq:Kamil-detail}
\begin{eqnarray}
K_{0} \!\! &=& \!\! \omega \bar L, \\
K_{1} \!\! &=& \!\! \frac{1}{2\pi}\int^{2\pi}_{0}H_{1}{\rm d} \bar l, \\
K_{2} \!\! &=& \!\! \frac{1}{2\pi}\int^{2\pi}_{0}
[2(H_{1};W_{1})+((H_{0};W_{1});W_{1})]{\rm d}\bar l, \\
K_{3} \!\! &=& \!\! \frac{1}{2\pi}\int^{2\pi}_{0}\Big[3(H_{1};W_{2})
+ 3((H_{1};W_{1});W_{1}) \nonumber \\
\!\! &+& \!\! 2((H_{0};W_{2});W_{1})
+ ((H_{0};W_{1});W_{2}) \nonumber \\
\!\! &+& \!\! (((H_{0};W_{1});W_{1});W_{1})\Big]{\rm d} \bar l.
\end{eqnarray}
\end{subequations}
$W_1$ and $W_2$ are determined through solving the following
differential equations
\begin{subequations}
\label{eq:W}
\begin{eqnarray}
\omega \frac{\partial W_{1}}{\partial \bar l} \!\!\! &=& \!\!
H_{1}(\bar l,\bar g,\bar p,\bar L,\bar G,\bar P) \!-\!
K_{1}(\bar g,\bar p,\bar L,\bar G,\bar P), \\
\omega \frac{\partial W_{2}}{\partial \bar l} \!\!\! &=& \!\!
2(H_{1};W_{1})+((H_{0};W_{1});W_{1})-K_{2}.
\end{eqnarray}
\end{subequations}
By substituting from (\ref{eq:W}) into (\ref{eq:Kamil-detail}) and
evaluating the integrals, one finds the explicit form of the new
Hamiltonian $K$, which has been given in Appendix \ref{sec:appA} up
to the third-order terms.

Once $\bar l$ is removed from the Hamiltonian, $\bar L$ becomes an
integral of motion. The slow dynamics of the system is thus governed
by the flows in the $(\bar g,\bar G)$-space. We introduce the {\it
slow time} $\tau=\bar p/\Omega$, ignore the fourth-order terms in
$\epsilon$, and obtain the following differential equations for the
dynamics of $(\bar g,\bar G)$
\begin{subequations}
\label{eq:norm-eq-of-motion}
\begin{eqnarray}
\frac{{\rm d} \bar g}{{\rm d} \tau} \!\! &=& \!\! \frac{\partial
K}{\partial \bar G} = f_{1}(\bar g,\bar G)+\epsilon
h_{1}(\bar g,\bar G,\epsilon,\tau), \\
\frac{{\rm d} \bar G}{{\rm d} \tau} \!\! &=& \!\! -\frac{\partial
K}{\partial \bar g} = f_{2}(\bar g,\bar G)+\epsilon h_{2}(\bar
g,\bar G,\epsilon,\tau),
\end{eqnarray}
\end{subequations}
where
\begin{eqnarray}
\label{eq:mel-comp} f_{1} \!\! &=& \!\! d_6+d_1 \cos (2\bar g),~~
f_{2} = e_1 \sin (2\bar g),\nonumber \\
h_{1} \!\! &=& \!\! d_7+d_2 \cos (2\bar g) \nonumber \\
\!\! &+& \!\! \epsilon [ d_8 + d_3 \cos (2\bar g) + d_4 \cos (4\bar
g) +d_5 \cos (2\bar g-2\Omega \tau)],\nonumber \\
h_{2} \!\! &=& \!\! e_2 \sin (2\bar g) \nonumber \\
\!\! &+& \!\! \epsilon [ e_3 \sin (2\bar g) + e_4 \sin (4\bar g) +
e_5 \sin (2\bar g-2\Omega \tau)].
\end{eqnarray}
In these equations, $d_i$ ($i=1,\cdots,8$) and $e_j$
($j=1,\cdots,5$) are functions of $\bar L$ and $\bar G$ 
(see Appendix \ref{sec:appA}). It is remarked that the
action $\bar P$ appears only in $K_1$ via the term 
$\Omega \bar P$. It then disappears in the normalized 
equations (\ref{eq:norm-eq-of-motion}) after taking the 
partial derivatives of $K$ with respect to $\bar g$ and $\bar G$.
The partial derivative of $K$ with respect to $\bar P$ 
determines the evolution of $\bar p$, which is in 
accordance with the simple linear law  
$\bar p(\tau)=\Omega \tau+\bar p(0)$. 
The dynamics of $\bar P$ itself is governed by
\begin{equation}
\frac {{\rm d}\bar P}{{\rm d}\tau}=
-\frac {\partial K}{\partial \bar p}=
-\frac{1}{\Omega} 
\frac {\partial}{\partial \tau}K(\bar g,\bar G,\tau).
\label{eq:dynamics-of-bar-P}
\end{equation}
One may integrate (\ref{eq:dynamics-of-bar-P}) to obtain $\bar P(\tau)$ 
once equations (\ref{eq:norm-eq-of-motion}) are solved. The behavior
of $\bar P$ is thus inherited from $\bar g(\tau)$ and $\bar G(\tau)$.

\section{The Melnikov Function}
\label{sec:melnikov} There are few analytical methods in the
literature for the detection of chaos in perturbed Hamiltonian
systems \cite{M63,Ch79}. Melnikov's \cite{M63} method is the most
powerful technique when the governing equations take the form
\begin{eqnarray}
\label{eq:mel-sys} \frac {{\rm d}\textbf{x}}{{\rm d}\tau} =
\textbf{f}(\textbf{x})
  +\epsilon\textbf{h}(\textbf{x},\epsilon,\tau),~~
\textbf{x}\in \mathbb{R}^{2},
\end{eqnarray}
so that the unperturbed system ${\rm d}\textbf{x}/{\rm d}\tau
=\textbf{f}(\textbf{x})$ is integrable and possesses a homoclinic
(heteroclinic) orbit $\textbf{q}_{\rm h}(\tau)$ to a hyperbolic
saddle point, and $\textbf{h}(\textbf{x},\epsilon,\tau)$ is
$T$-periodic in $\tau$. The occurrence of chaos is examined by the
Melnikov function
\begin{eqnarray}
\label{eq:mel-fun-general} M(\tau_{0},\epsilon) = \epsilon M_{1}(\tau_{0})
+ \epsilon^{2} M_{2}(\tau_{0})+ \ldots,
\end{eqnarray}
where $M_{k}(\tau_{0})$ denotes the $k$th-order Melnikov function.
Assume that $M_{i}(\tau_{0})$ is the first nonzero term, i.e.,
$M_{k}(\tau_{0})\equiv 0$ for $1\leq k \leq i-1$. If $M_{i}(\tau_0)$
has simple zeros, then, for sufficiently small $\epsilon$, the
system (\ref{eq:mel-sys}) has transverse homoclinic (heteroclinic)
orbits, which imply chaos due to the Smale-Birkhoff homoclinic
theorem \cite{GH83}. The first-order term in
(\ref{eq:mel-fun-general}) is determined by the classical formula
\begin{eqnarray}
\label{eq:mel-fun} M_{1}(\tau_{0}) = \int^{+\infty}_{-\infty}
{\bf{f}}\left({\bf{q}}_{\rm h}\left(\tau\right)\right) \wedge
{\bf{h}}\left({\bf{q}}_{\rm h}\left(\tau\right),0,\tau + \tau_{0}
\right) \rm{d}\tau,
\end{eqnarray}
where the wedge operator $\wedge$ is defined as $\textbf{f}\wedge
\textbf{h} = f_1 h_2-f_2 h_1$. Although the Hamiltonian equations
(\ref{eq:norm-eq-of-motion}) have a suitable form for the 
application of Melnikov's method, they are autonomous up to the 
first-order terms in $\epsilon$. Consequently, $M_1(\tau_{0})$ 
vanishes identically for all $\tau_0\in[0,T]$. We thus need to 
investigate the second-order Melnikov function. For doing so, 
we begin with solving the unperturbed system   
\begin{subequations}
\label{eq:unper-eq-of-motion}
\begin{eqnarray}
\frac{{\rm d} \bar g}{{\rm d} \tau} &=&
\frac{\partial K_{1}}{\partial \bar G} = f_{1}(\bar g,\bar G), \label{eq:unper-g}\\
\frac{{\rm d} \bar G}{{\rm d} \tau} &=& -\frac{\partial
K_{1}}{\partial \bar g} = f_{2}(\bar g,\bar G),\label{eq:unper-G}
\end{eqnarray}
\end{subequations}
along homoclinic (heteroclinic) orbits. 
Jalali and Angoshtari \cite{JA1} showed that for 
$\bar L>\eta\omega^3 /\gamma$, equations (\ref{eq:unper-eq-of-motion})
have hyperbolic stationary points at $S_0\equiv (\bar g_0,\bar
G_0)=(-\pi,0)$, $S_2\equiv(\bar g_2,\bar G_2)=(0,0)$, and
$S_4\equiv(\bar g_4,\bar G_4)=(\pi,0)$. The {\it implicit} equation
of the invariant manifolds that terminate at the saddle points are
\begin{equation}
\label{eq:homo-heter} \cos [2\bar g_{\rm h}(\tau)] = \left [\bar
L-\frac{\gamma}{2\eta\omega^3}\bar G_{\rm h}^2(\tau) \right]
\left[\bar L^2-\bar G_{\rm h}^2(\tau)\right]^{-1/2}.
\end{equation}
\begin{figure*}[t]
\centerline{\hbox{\includegraphics[width=0.29\textwidth]{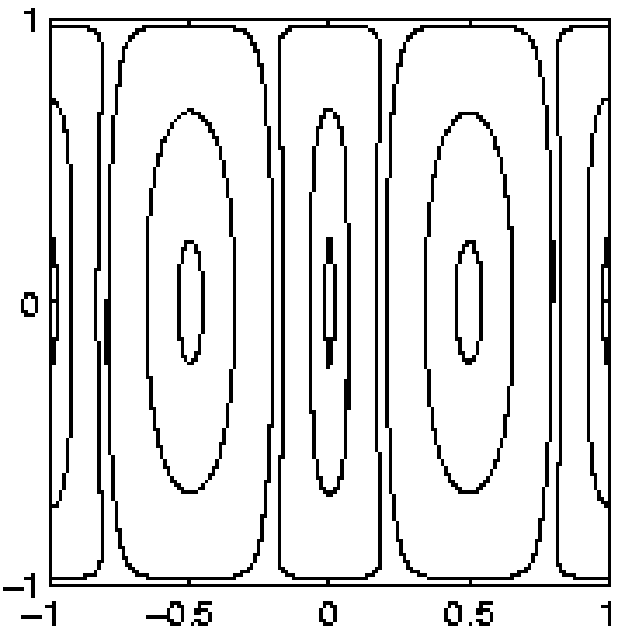}
                  \hspace{0.1cm}
                  \includegraphics[width=0.3\textwidth]{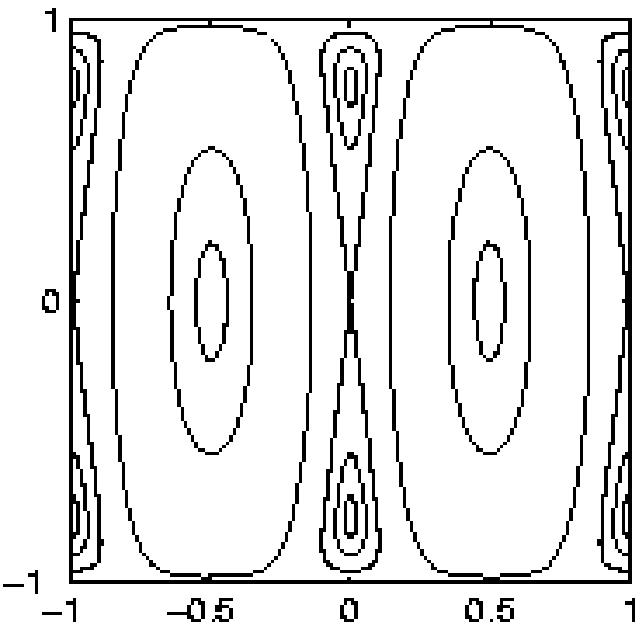}
                  \hspace{0.1cm}
                  \includegraphics[width=0.29\textwidth]{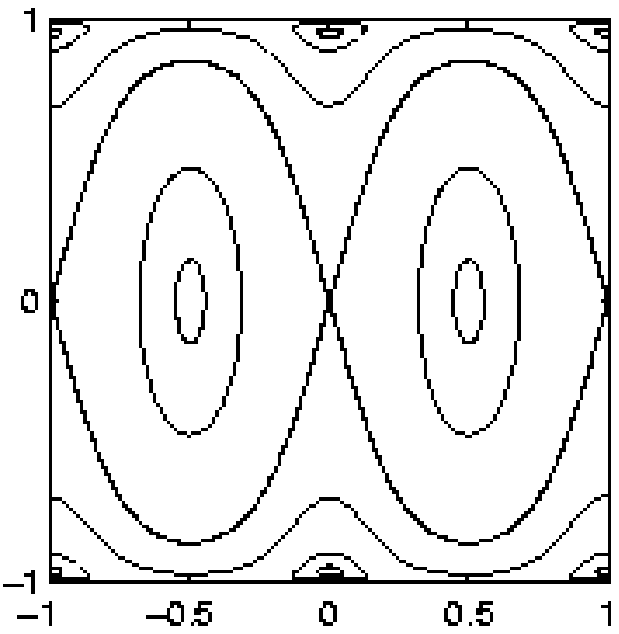}}}
\caption{Possible topologies of the phase space flows of the
averaged system for $F=0$. In all panels the horizontal axis indicates
$\bar g/\pi$ and the vertical axis indicates $\bar G/\bar L$. In the
first topology (left panel), all stationary points are of center type.
In the second topology (middle panel) two new centers, with non-zero
$\bar G$-coordinates, have emerged for $\bar g=\pm n\pi$ ($n=0,1$)
and symmetrical homoclinic loops (thick lines) connect saddle points
to themselves. In the third topology (right panel) the off-axis centers
(and their surrounding tori) are still present but the separatrix curves
(thick lines) are of heteroclinic type. \label{fig:fig1}}
\end{figure*}
\begin{figure*}[]
\centerline{\hbox{\includegraphics[width=0.47\textwidth]{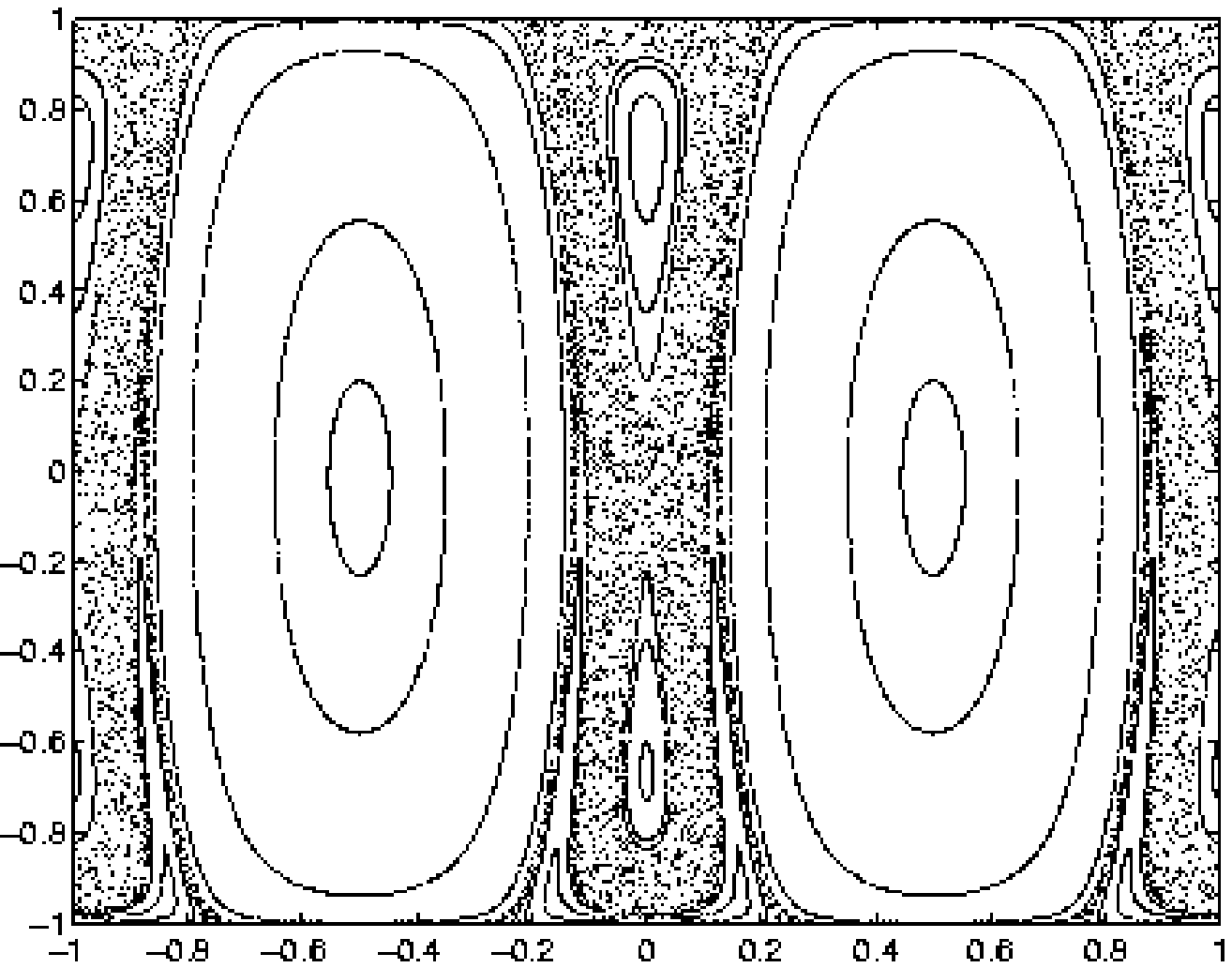}
                  \hspace{0.15cm}
                  \includegraphics[width=0.47\textwidth]{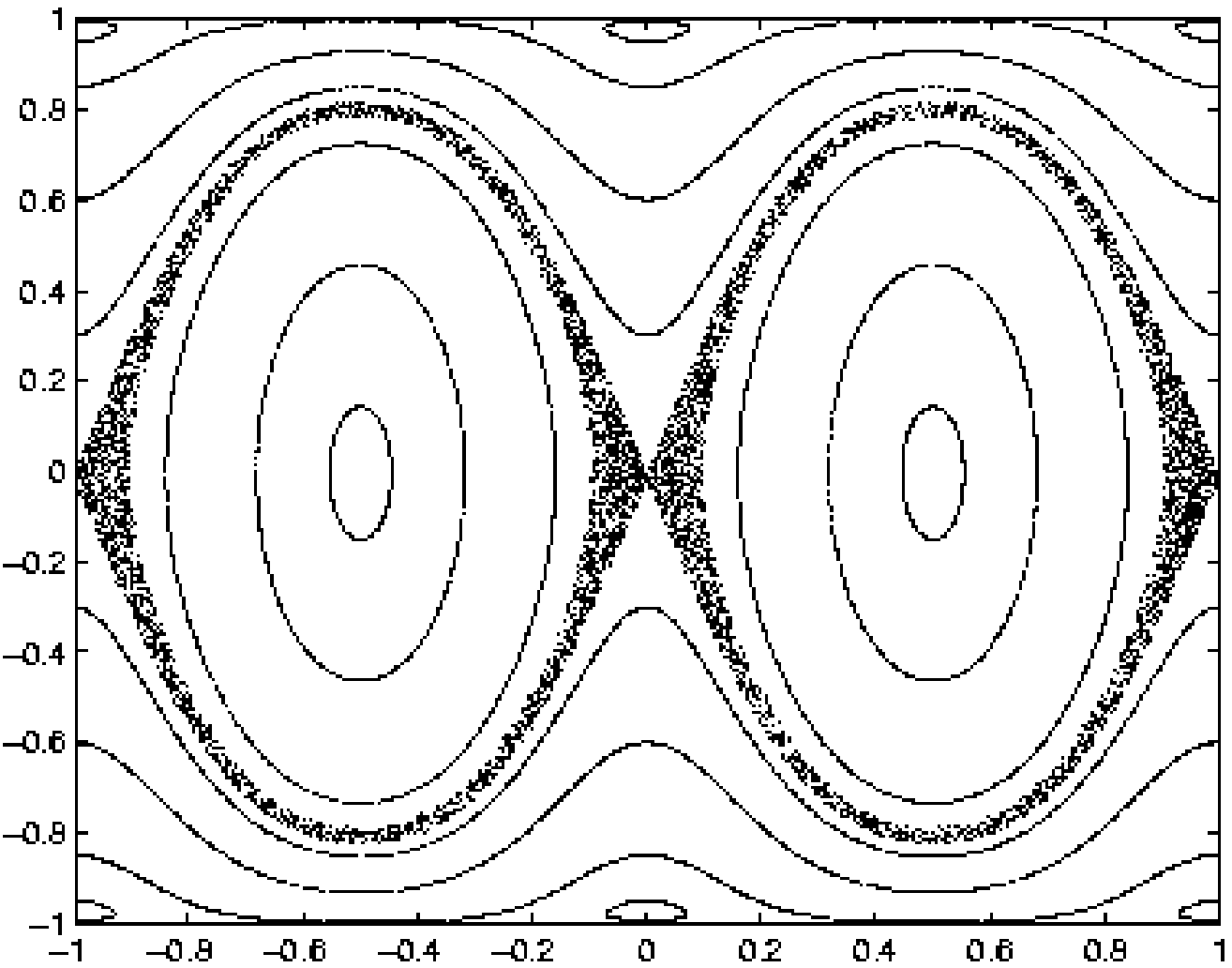}}}
\caption{Phase space structure of the normalized equations
(\ref{eq:norm-eq-of-motion}) with $\bar L=2$, $\omega=\eta=F=1$, and
$\Omega=\epsilon=0.1$. {\it Left panel}: $\gamma=0.8$. {\it Right
panel}: $\gamma=2$. In both panels the horizontal and vertical axes
indicate $\bar g/\pi$ and $\bar G/\bar L$, respectively. \label{fig:fig2}}
\end{figure*}

For $\gamma \bar L\ge 2\eta\omega^3$, equation (\ref{eq:homo-heter})
represents a heteroclinic orbit which connects $S_0$ to $S_2$. For
$\eta\omega^3 < \gamma \bar L < 2\eta\omega^3$, the heteroclinic
orbit disappears and it is replaced by a homoclinic orbit
(see Figure \ref{fig:fig1}). To compute the {\it explicit} form of
the homoclinic (or heteroclinic) orbit of (\ref{eq:unper-eq-of-motion}),
we use (\ref{eq:unper-G}) and (\ref{eq:homo-heter}), and obtain
\begin{eqnarray}
&{}& \int^{\bar G_{\rm h}(\tau)}_{\bar G_{\rm h}(0)} \frac{{\rm d}
\bar G}{\bar G \sqrt{1-\alpha \bar G^2}} = \pm \sqrt{\beta} \tau,
\label{eq:int-G-q0-tau} \\
&{}& \alpha = \frac{\gamma^2}{16\eta \beta \omega^4},~~ \beta =
\frac{\gamma \bar L - \eta\omega^3}{4\omega}, \nonumber
\end{eqnarray}
where the lower integration limit is $\bar G_{\rm
h}(0)=1/\sqrt{\alpha}$. After taking the integral
(\ref{eq:int-G-q0-tau}), we arrive at
\begin{eqnarray}
\label{eq:G-tau} \bar G_{\rm h}(\tau) &=& \frac{\rm sech
(\sqrt{\beta}\:\tau)}{\sqrt{\alpha}},
\end{eqnarray}
for the $\bar G\ge 0$ branch of the homoclinic (heteroclinic)
orbit. Having $\bar G_{\rm h}(\tau)$, it is straightforward to 
calculate $\cos[2\bar g_{\rm h}(\tau)]$ and 
$\sin[2\bar g_{\rm h}(\tau)]$, and determine the explicit form of 
$\mathbf{q}_{\rm h}(\tau)=
\left [\bar g_{\rm h}(\tau),\bar G_{\rm h}(\tau)\right ]$.  

For constructing $M_2(\tau_0)$, we use Fran\c{c}oise's \cite{FR96,PRK01} 
algorithm that has been devised for dynamical systems with polynomial 
nonlinearities. To express the averaged Hamiltonian $K$ in terms of 
polynomial functions of some new dependent variables, we utilize 
Hopf's variables  
\begin{subequations}
\begin{eqnarray}
\label{eq:hopf}
Q_1 &=& \frac 1{2\omega} \sqrt{\bar L^2-\bar G^2} \cos (2\bar g), \\
Q_2 &=& \frac 1{2\omega} \sqrt{\bar L^2-\bar G^2} \sin (2\bar g),
\end{eqnarray}
\end{subequations}
and obtain the following differential 1-form for the evolution
of the averaged system
\begin{eqnarray}
&{}& {\partial K_1\over \partial Q_1}{\rm d}Q_1 + {\partial K_1\over
\partial Q_2}{\rm d}Q_2 \nonumber \\
&+& \epsilon \left[\left( s_1 +\epsilon s_2\right){\rm d}Q_2 -\left(
z_1 +\epsilon z_2\right){\rm d}Q_1\right]=0.
\end{eqnarray}
Here, the first-order Hamiltonian is
\begin{equation}
\label{eq:K1-Q}
K_1 (Q_1 , Q_2) = \frac{\gamma}{2}(Q^{2}_{1}+Q^{2}_{2})
-\frac{\eta \omega^2}{2}Q_1+C,
\end{equation}
and
\begin{subequations}
\label{eq:detail-dif-form}
\begin{eqnarray}
s_1 &=& m_1 Q_2, \\
z_1 &=& n_1 Q_1 + n_6,  \\
s_2 &=& m_2 (Q^{2}_{1}+Q^{2}_{2})Q_{2} + m_3 Q_{1}Q_{2}
+m_4 Q_{2} \nonumber \\ &+& m_5 \sin (2\Omega \tau), \\
z_2 &=& n_2 (Q^{2}_{1}+Q^{2}_{2})Q_{1} + n_3 (3Q^{2}_{1}+Q^{2}_{2})
+n_4 Q_{1} \nonumber \\ &+& n_5 \cos (2\Omega \tau) +n_7.
\end{eqnarray}
\end{subequations}
The constant coefficients $C$, $m_i$ ($i=1,\cdots,5$), and $n_j$
($j=1,\cdots,7$) have been given in Appendix \ref{sec:appB}. A
prerequisite for the application of Fran\c{c}oise's \cite{FR96}
algorithm is that for all polynomial 1-forms $D$ that satisfy
the condition
\begin{equation}
\int_{\textbf{q}_{\rm h}}D\equiv 0,
\end{equation}
there must exist polynomials $A(Q_1,Q_2)$ and $r(Q_1 , Q_2)$ such
that $D={\rm d}A+r{\rm d}K_1$. We call this the condition $(*)$
and prove in Appendix \ref{sec:appC} that $K_1$ satisfies the
condition $(*)$.

Fran\c{c}oise's algorithm states that if $M_1(\tau_0)=\cdots=M_{k-1}
(\tau_0)\equiv 0$ for some integer $k\geq 2$, it follows that
\begin{subequations}
\label{eq:M_k}
\begin{eqnarray}
M_k (\tau_0) &=& \int_{\textbf{q}_{\rm h}}D_k, \\
D_1 &=& \delta_1 ,\ D_m = \delta_m +\sum_{i+j=m} r_i \delta_j , \\
\delta_j &=& z_j {\rm d}Q_1 - s_j {\rm d}Q_2, \label{eq:M_k_c}
\end{eqnarray}
\end{subequations}
for $2\leq m\leq k$. The functions $r_i$ are then determined
successively from the formulas $D_i={\rm d}A_i +r_i{\rm d}K_1$
for $i=1,\cdots,k-1$. We have already found that
\begin{eqnarray}
M_1 (\tau_0) &=& \int_{\textbf{q}_{\rm h}}\delta_1=0, \\
\delta_1 &=& (n_1 Q_1 +n_6 ){\rm d}Q_1 -m_1 Q_2 {\rm d}Q_2.
\label{eq:omega1}
\end{eqnarray}
From (\ref{eq:M_k}) and (\ref{eq:Omega-d-1}) it can be shown that
\begin{equation}
\label{eq:M-2}
M_2 (\tau_0)=\int_{\textbf{q}_{\rm h}}\delta_2.
\end{equation}
Substituting (\ref{eq:M_k_c}) and (\ref{eq:detail-dif-form})
into (\ref{eq:M-2}), and carrying out the integration along 
$\textbf{q}_{\rm h}(\tau)$, result in
\begin{eqnarray}
M_{2}(\tau_{0}) = \frac{3\gamma F^2 I}{2\omega^6} \sin
(2\Omega\tau_{0}), \label{eq:mel-final}
\end{eqnarray}
with $I$ being a constant (see Appendix \ref{sec:appD}). Equation
(\ref{eq:mel-final}) shows that $\tau_n=n\pi/(2\Omega)$ ($n=
1,2,\cdots$) are simple zeros of $M_2 (\tau_0)$ so that
\begin{equation}
M_2 (\tau_{n})=0,~~ \frac{\partial M_2 (\tau_0)}{\partial \tau_0}
\Big\vert _{\tau_0=\tau_n} \neq 0. \label{eq:condition}
\end{equation}
Thus, we conclude that the global stable and unstable manifolds of
the saddle point $S^{\tau_0}_n$, $W^{s}(S^{\tau_0}_{n})$ and
$W^{u}(S^{\tau_0}_{n})$, always intersect transversely. Transversal
intersections cause a sensitive dependence on initial conditions due
to the Smale-Birkhoff homoclinic theorem. This is a route to chaos.
On the other hand this means that the reduced equations
(\ref{eq:norm-eq-of-motion}) are non-integrable for $F\not =0$.

\section{Classification of circumferential waves}
\label{sec:wave-classification} For $F=0$, $\textbf{h}$ does not
depend on $\tau$ and the normalized equations
(\ref{eq:norm-eq-of-motion}) are integrable. In such a circumstance,
the phase space structure can take three general topologies
(depending on the values of the system parameters and $\bar L$) as
shown in Figure \ref{fig:fig1}.
In the first topology all stationary points with
the coordinates $(\bar g_s,\bar G_s)$ calculated from
\begin{equation}
\textbf{f}(\bar g_s,\bar G_s)+\epsilon \textbf{h}(\bar g_s,\bar
G_s,\epsilon)=\textbf{0},
\end{equation}
are centers and they lie on the $\bar G=0$ axis with $\bar
g_s=-\pi+n\pi/2$ ($n=0,\cdots,4$). In the second and third
topologies, two off-axis centers (with $\bar G_s\not =0$) come to
existence for $\bar g_s=\pm n\pi$ ($n=0,1$) and the on-axis
stationary points with the same $\bar g_s=\pm n\pi$ become saddles.
In the second topology, each saddle point is connected to itself by
a homoclinic orbit, and in the third topology, a heteroclinic orbit
connects two neighboring saddle points. The system with heteroclinic
orbits allows for rotational $\bar g(\tau)$ while in the system with
homoclinic orbits $\bar g(\tau)$ is always librating. Beware that
this classification of phase space flows is valid as long as
$\epsilon$ is sufficiently small.

For $F=0$, the phase space flows of (\ref{eq:norm-eq-of-motion}) are
structurally stable (with no unbounded branches) and the whole
$(\bar g,\bar G)$-space is occupied by periodic orbits of period
$T(K)$. At the stationary points, one has $T(K(\bar g_s,\bar
G_s,\bar L))=0$. Given the invariance of $\bar L$, and the periodic
solutions $\bar g(\tau)=\bar g(\tau+T(K))$ and $\bar G(\tau)=\bar
G(\tau+T(K))$, the anomaly $\bar l$ is determined through solving
\begin{equation}
{{\rm d}\bar l \over {\rm d}t}=\omega + \epsilon {\partial \over
\partial \bar L}\left ( K_1+{\epsilon \over 2!}
K_2+{\epsilon ^2\over 3!}K_3 \right ),
\end{equation}
which results in $\bar l=\omega t +\epsilon R(\tau)$ with
$R(\tau)=R(\tau+T(K))$. According to (\ref{eq:Pri-unPri}), the
functions $g(t)$, $G(t)$ and $L(t)$ are also periodic in $t$ and we
conclude that $l(t)=\omega t+R_W(t)$ with $R_W(t)=E_W(\bar l)-\omega t$
being a small-amplitude periodic function of $t$. The explicit from
of the circumferential wave will then become
\begin{eqnarray}
\!\!\! &{}& \!\!\! {w(R,\phi,t) \over U_m(R)} =
\sqrt{L(t)\!+\!G(t)\over 2\omega}
\cos\left [n\phi \!-\! \omega t \!-\! R_W(t)\!-\! g(t) \right ] \nonumber \\
\!\! &{}& \!\! \qquad -\sqrt{L(t) \!-\! G(t)\over 2\omega} \cos\left
[n\phi \!+\! \omega t \!+\! R_W(t) \!-\! g(t) \right ],
\label{eq:wave-lissajous}
\end{eqnarray}
which is composed of a forward and a backward traveling wave. Due to
the periodic nature of $L(t)$ and $G(t)$, when the amplitude of the
forward traveling wave is maximum, that of the backward wave is
minimum and vice versa.

As our results of \S\ref{sec:melnikov} shows, the regular nature of
traveling waves is destroyed for $F\not =0$ and a chaotic layer
occurs through the destruction of the homoclinic and heteroclinic
orbits of (\ref{eq:unper-eq-of-motion}). This happens over the time
scale $\tau \sim {\cal O}(\epsilon^{-2})$ or $t\sim {\cal
O}(\epsilon^{-3})$ (because $\bar p$ is present only in $K_3$).
Figure \ref{fig:fig2} shows Poincar\'e maps of the system
(\ref{eq:norm-eq-of-motion}) for $F\not =0$. The sampling time step
in generating the Poincar\'e maps has been $2\pi/\Omega$. It is seen
that most tori around elliptic fixed points are preserved. They
correspond to regular periodic and quasi-periodic solutions of the
normalized system. For chaotic flows, the functions $\bar g(\tau)$
and $\bar G(\tau)$ randomly change within the invariant measure of
the chaotic set. Consequently, the original Lissajous variables
$g(t)$, $L(t)$, $G(t)$ and also $R_W(t)$ become chaotic too.

For $\bar G(\tau)>0$ and $\bar G(\tau)<0$ the forward and the
backward traveling waves are the dominant components of the
circumferential wave, respectively. When the chaotic layer emerges
from the destroyed homoclinic orbits (left panel in
Figure \ref{fig:fig2}), the
sign of $\bar G(\tau)$ is randomly switched along a chaotic
trajectory. This means a random transfer of kinetic/potential energy
between the forward and backward traveling wave components. For
chaotic trajectories of this kind the angle $\bar g(\tau)$ randomly
fluctuates near $\bar g\approx \pm n\pi$ ($n=0,1$) with an almost
zero average. The evolution of circumferential waves is quite
different when the chaotic layer emerges due to destroyed
heteroclinic orbits (right panel in Figure \ref{fig:fig2}).
In this case chaos
means a random change between the librational and rotational states
of $\bar g(\tau)$. Such a change induces an unpredictable phase
shift of magnitude $\pi$ for both forward and backward traveling
wave components. We note that $\bar G(\tau)$ can flip sign on a
chaotic trajectory only when $\bar g(\tau)$ is in its librational
state.

\section{concluding remarks}
\label{sec:conclusions}

Resonance overlapping \cite{Ch79,Cont02} is the main cause for
chaotic behavior in spinning disks with near-resonant angular
velocities \cite{RM01,JA1}. The chaos predicted in this paper,
however, happens far from fundamental resonances. Optical and HHDs
are usually operated below critical resonant speeds and the lateral
force $F$ due to magnetic head is very small. We showed that
whatever the magnitude of $F$ may be, a chaotic layer fills some
parts of the phase space because the Melnikov function of the
normalized equations has {\it always} simple zeros. Dynamics of
rotating disks is regular only if $F$ vanishes, which is an
unrealistic assumption for disk drives. In low-speed disks with
small $F$, diffusion of chaotic orbits (within their invariant
measure) takes a long time of $t\sim {\cal O}(\epsilon^{-3})$. The
slow development of chaotic circumferential waves makes them
undetectable in short time scales at which most controllers work.
The Melnikov function (\ref{eq:mel-final}) depends not only on $F$,
but also on the parameter $\eta$ through the constant $I$. The
parameter $\eta$ is a contribution of imperfections, which are
likely because of limited fabrication precision in micro/nano
scales. For a perfect disk with $\eta=0$, the off-axis elliptic
stationary points of (\ref{eq:unper-eq-of-motion}), and
consequently, homoclinic and heteroclinic orbits disappear. In such
a condition the Melnikov function is indefinite, but the system
admits an exact first integral and the dynamics is governed by the
Hamiltonian function given in equation (11) of Jalali and Angoshtari
\cite{JA1}.

One of the most important achievements of this work was to unveil
the fact that it is premature to truncate the series of canonical
perturbation theories before recording the role of all participating
variables. In systems with non-autonomous governing ODEs
(non-conservative systems), one must be cautious while removing a
fast angle through averaging schemes. The removal of the fast angle
may also wipe out time-dependent terms, up to some finite orders of
$\epsilon$, and hide some essential information of the underlying
dynamical process. Strange irregular solutions can indeed occur at
any order and influence the long term response of dynamical systems
as we observed for the spinning disk problem by keeping the
third-order terms.

\acknowledgments
We are indebted to the anonymous referee, who discovered an error
in the early version of the paper and led us to investigate 
the second-order Melnikov function. MAJ thanks the Research 
Vice-Presidency at Sharif University of Technology for 
partial support.

\appendix
\section{The normalized Hamiltonian}
\label{sec:appA} By evaluating the integrals in (\ref{eq:Kamil-detail}),
we obtain the first, second and third order terms of the normalized
Hamiltonian as
\begin{eqnarray}
\label{eq:K-1-2-3} K_{1} &=& \frac{-\gamma \bar G^2}{8\omega^2}+
\frac{{\bar L}(3\gamma {\bar L}+2\eta \omega^3 )}{8\omega^2}+\Omega {\bar P} \nonumber \\
&-& \frac{\eta \omega}{4}\sqrt{\bar L^2 - \bar G^2} \cos (2\bar g), \\
K_{2} &=& -\frac{1}{64\omega^5}\Big [2\bar L(-9 \bar G^{2}+17 \bar
L^2 )\gamma^2+
64 F^{2} \omega^3 \nonumber \\
&+& 8(3 \bar L^2-\bar G^2) \gamma\eta\omega^3 +8\bar L\eta^2 \omega^6
\nonumber \\
&+& 4\omega^3 [-6\bar L\gamma\eta -2\eta^2 \omega^3]\sqrt{\bar L^2 -
\bar G^2}\cos (2\bar g)\Big], \\
K_{3} &=& \frac{3}{512\omega^8}\Big \{11 \bar G^4 \gamma^3 -258 \bar
G^2 \bar L^2\gamma^3 + 375 {\bar L}^4 \gamma^3 \nonumber \\ &+& 1024
F^2 {\bar L}\gamma\omega^3 -180 {\bar G}^2 {\bar L}\gamma^2
\eta\omega^3
+ 340 {\bar L}^3 \gamma^2 \eta\omega^3  \nonumber \\
&+& 256F^2 \eta\omega^6 -48{\bar G}^2 \gamma\eta^2 \omega^6 +
176{\bar L}^2 \gamma\eta^2 \omega^6 \nonumber \\
&+& 32{\bar L}\eta^3 \omega^9 -
2\omega^3 \Big [ 17(10{\bar L}^2 -{\bar G}^2 ) \gamma^2 \eta \nonumber \\
&+& 96{\bar L}\gamma\eta^2 \omega^3 +16\eta^3 \omega^6 \Big ]
\sqrt{{\bar L}^2 - {\bar G}^2}\cos (2{\bar g}) \nonumber \\
&-& 16({\bar G}^2 -{\bar L}^2)\gamma\eta^2 \omega^6 \cos (4{\bar
g})+256F^2 \eta\omega^6 \cos (2{\bar p})\nonumber \\
&-& 512F^2 \gamma\omega^3 \sqrt{{\bar L}^2-{\bar G}^2}\cos (2{\bar
g}-2{\bar p})\Big \}.
\end{eqnarray}
Consequently, the functions $d_i(\bar L,\bar G)$ in equations
(\ref{eq:mel-comp}) are found to be
\begin{eqnarray}
\label{eq:mel-comp-detail1}
d_{1} &=& \frac{\eta\omega {\bar G}}{4} \left({\bar L}^2 -
{\bar G}^2 \right)^{-1/2}, \nonumber \\
d_{2} &=& -\frac{{\bar G}}{16\omega^2}(6{\bar L}\gamma\eta+
2\eta^2 \omega^3 )({\bar L}^2 - {\bar G}^2)^{-1/2}, \nonumber \\
d_{3} &=& \frac{3{\bar G}}{256\omega^5} \Big [16\eta^3 \omega^6 +
51(4{\bar L}^2-{\bar G}^2)\gamma^2 \eta \nonumber \\
&+& 96{\bar L}\gamma \eta^2 \omega^3 \Big ]({\bar L}^2 -
{\bar G}^2)^{-1/2}, \nonumber \\
d_{4} &=& -\frac{3{\bar G}\gamma \eta^2}{16\omega^2},\nonumber \\
d_{5} &=& \frac{3{\bar G} F^2 \gamma}{\omega^5}
({\bar L}^2 - {\bar G}^2)^{-1/2}, \nonumber \\
d_{6} &=& -\frac{\gamma {\bar G}}{4\omega^2},\nonumber \\
d_{7} &=& \frac{{\bar G}}{32\omega^5}(18{\bar L}\gamma^2+
8\gamma\eta\omega^3 ), \nonumber \\
d_{8} &=& \frac{3{\bar G}}{256\omega^8}\Big [2\gamma^3
(11{\bar G}^2 - 129{\bar L}^2) - 180{\bar L}\gamma^2 \eta \omega^3 \nonumber \\
&-& 48\gamma \eta^2 \omega^6 \Big ].
\end{eqnarray}
Defining $S=(\bar L^2-\bar G^2)/{\bar G}$, the functions $e_j(\bar
L,\bar G)$ in equations (\ref{eq:mel-comp}) become
\begin{eqnarray}
\label{eq:mel-comp-detail2}
e_{j} &=& -2d_{j}S, \ \ j=1,\ldots,5,\:j\neq3, \nonumber \\
e_{3} &=& -\frac{3}{128\omega^5}\Big [ 16\eta^3 \omega^6 +
17(10{\bar L}^2-{\bar G}^2) \gamma^2 \eta \nonumber \\
&+& 96{\bar L}\gamma \eta^2 \omega^3 \Big]\sqrt{{\bar L}^2 -{\bar
G}^2}.
\end{eqnarray}

\section{}
\label{sec:appB}
The constant coefficients of equations (\ref{eq:K1-Q}) and
(\ref{eq:detail-dif-form}) are as follows
\begin{eqnarray}
\label{eq:C-m-n}
C &=& \frac{{\bar L}(\gamma {\bar L}+\eta\omega^3)}{4\omega^2}, \nonumber \\
n_1 &=& \frac{1}{8\omega^3}(9{\bar L}\gamma^2+4\gamma\eta\omega^3), \nonumber \\
n_2 &=& -\frac{33\gamma^3}{48\omega^4}, \nonumber \\
n_3 &=& \frac{17\gamma^2 \eta}{64\omega^2}, \nonumber \\
n_4 &=& -\frac{1}{32\omega^6}(59{\bar L}^2 \gamma^3 +45{\bar L}\gamma^2 \eta\omega^3
 +16\gamma\eta^2 \omega^6), \nonumber \\
n_5 &=& \frac{F^2 \gamma}{\omega^4}, \nonumber \\
n_6 &=& -\frac{3{\bar L}\gamma\eta+\eta^2 \omega^3}{8\omega}, \nonumber \\
n_7 &=& \frac{1}{256\omega^4}(153{\bar L}^2 \gamma^2 \eta +96{\bar L}\gamma\eta^2 \omega^3
 +16\eta^3 \omega^6), \nonumber \\
m_i &=& -n_i , \ \ i=1,2,5, \nonumber \\
m_3 &=& -2n_3 , \nonumber \\
m_4 &=& \frac{1}{32\omega^6}(59{\bar L}^2 \gamma^3 +45{\bar L}\gamma^2 \eta\omega^3
 +8\gamma\eta^2 \omega^6).\nonumber \\
\end{eqnarray}

\section{}
\label{sec:appC} In this appendix we prove that $K_1(Q_1,Q_2)$ given
in (\ref{eq:K1-Q}), satisfies the condition $(*)$. To this end, we
need the following theorem.

\textbf{Theorem 1.} \textit{Any polynomial 1-form $D$ of degree
$n$ in $Q_1$ and $Q_2$ can be expressed as
\begin{eqnarray}
\label{eq:omega-decomp}
D = {\rm d}A +r{\rm d}K_1 +\xi(K_1 )Q_2 {\rm d}Q_1 ,
\end{eqnarray}
where $A(Q_1 ,Q_2 )$ and $r(Q_1 ,Q_2 )$ are polynomials of degree
$(n+1)$ and $(n-1)$ respectively, and $\xi(K_1)$ is a polynomial
of degree $[\frac{1}{2} (n-1)]$ where $[x]$ denotes the greatest
integer in $x$.}

Iliev \cite{IL99} has proved the same theorem for
$H=(Q^{2}_{1}+Q^{2}_{2})/2$. Theorem 1 can thus be proved in a
similar manner. Here we only present a useful result.

Let $D$ be a general polynomial 1-form of degree 1,
\begin{subequations}
\label{eq:Omega-d-1}
\begin{eqnarray}
D = (a_{10} Q_1 +a_{01} Q_2 +a_{00} ){\rm d}Q_1 \nonumber \\
+ (b_{10} Q_1 +b_{01} Q_2 +b_{00} ){\rm d}Q_2 ,
\end{eqnarray}
then in (\ref{eq:omega-decomp}) we have
\begin{eqnarray}
A(Q_1 ,Q_2 ) &=& \frac{a_{10}}{2}Q^{2}_{1}+b_{10} Q_1 Q_2+\frac{b_{01}}{2}Q^{2}_{2} \nonumber \\
&+& a_{00} Q_1 +b_{00} Q_2,\\
r(Q_1 ,Q_2 ) &=& 0,\\
\xi (K_1 ) &=& a_{01} -b_{10}.
\end{eqnarray}
\end{subequations}
Since ${\rm d}K_1 =0$ along any phase space orbit characterized by
$K_1(Q_1,Q_2)=k$, and since the integral of an exact differential
${\rm d}A$ around any closed curve is zero, from
(\ref{eq:omega-decomp}) we obtain
\begin{eqnarray}
\int_{\textbf{q}_{\rm h}}D &=&
\xi(k)\int_{\textbf{q}_{\rm h}}Q_2{\rm d}Q_1 \nonumber \\
&=&\xi(k)\int^{+\infty}_{-\infty}Q_2(\tau)
\frac{{\rm d}Q_1}{{\rm d}\tau} {\rm d}\tau. \nonumber
\end{eqnarray}
On the other hand, from (\ref{eq:hopf}) we have
\begin{eqnarray}
\frac{{\rm d}Q_1}{{\rm d}\tau}=E(\tau)-2Q_2,\nonumber
\end{eqnarray}
where $E(\tau)$ is an even function of $\tau$. Given the fact that
$Q_2$ is an odd function of $\tau$, we conclude that
\begin{eqnarray}
\int_{\textbf{q}_{\rm h}}D = -2\xi(k)\int^{+\infty}_{-\infty}Q^2_2 {\rm d}\tau.\nonumber
\end{eqnarray}
Consequently, if $\int_{\textbf{q}_{\rm h}}D\equiv 0$, it follows
that $\xi(k)\equiv 0$ and therefore $D = {\rm d}A +r{\rm d}K_1$,
which completes the proof.

\section{}
\label{sec:appD} In equation (\ref{eq:mel-final}), the constant
coefficient $I$ is
\begin{eqnarray}
\label{eq:c-I} I &=& -\frac{64\pi\sqrt{\eta}\Omega
^2\omega^2}{\gamma} {\rm csch}\left(\frac{\pi\Omega}{\sqrt{\beta}}\right) \nonumber \\
&+& \frac{\pi\omega \sqrt{\eta}(3\gamma {\bar L}
-4\eta\omega^3)(\beta+4\Omega^2)}{\gamma \beta} {\rm
sech}\left(\frac{\pi\Omega}{\sqrt{\beta}}\right ) \nonumber \\
&-& \frac{ \pi\omega {\bar L}\sqrt{\eta}(3\beta-4\Omega^2)}{\beta}
{\rm sech}\left(\frac{\pi\Omega}{\sqrt{\beta}}\right) \nonumber \\
&+& \frac{4\pi\eta^{3/2}\omega^4}{\gamma} {\rm
sech}\left(\frac{\pi\Omega}{\sqrt{\beta}}\right ). \nonumber
\end{eqnarray}

\end{document}